\documentclass{article}
\usepackage{amsmath, amsthm}
\usepackage{amsfonts}

\newtheorem{theorem}{Theorem}[section]
\newtheorem{lemma}[theorem]{Lemma}

\begin{document}
\title{Gauge-invariant ground state for canonically quantized Yang-Mills theory}
\author{Rachel Lash Maitra}
\maketitle
\begin{abstract}
We use Hamilton-Jacobi theory to construct a gauge-invariant zero-energy candidate ground state for canonically quantized Yang-Mills theory with a ``nonlinear normal'' factor ordering, generalizing an analogous ordering introduced by Moncrief and Ryan for problems with finitely many degrees of freedom.  Invariance under spatial rotations and translations is immediate; boost invariance remains under investigation.  The motivation is to find a model for constructing a candidate ground state in general relativity, canonically quantized a la the Ashtekar variables.  We seek to avoid replicating some of the more troublesome features of the Kodama state, inherited from the Chern-Simons state.
\end{abstract}

\section{Introduction}
\label{Intro}

This paper presents a factor ordering of the canonically quantized
Yang-Mills Hamiltonian operator, and a corresponding gauge-invariant
candidate ground state in what might be called the Schr\"{o}dinger
representation. \ As usual in a canonically quantized gauge theory, the
``position'' variable is the vector potential $A$ of the gauge connection. \
The corresponding momentum $E$ is given by the (negative) Yang-Mills
electric field variable. \ These variables satisfy the Poisson bracket
relations
\begin{eqnarray*}
\Bigl\{ A_{i}^{I}\left( x\right) ,A_{j}^{J}\left( y\right) \Bigr\}
 = 0 = \left\{ E_{I}^{i}\left( x\right), E_{J}^{j}\left( y\right) \right\}, 
 \    \ 
\left\{ E_{J}^{j}\left( y\right) ,A_{i}^{I}\left( x\right)
\right\} =\delta ^{3}\left( x,y\right) \delta _{i}^{j}\delta _{J}^{I}
\end{eqnarray*}
and can be promoted to quantum operators as%
\begin{equation}
\begin{array}{l}
\hat{A}_{i}^{I}(x):\psi (A)\rightarrow A_{i}^{I}(x)\psi (A) \\ 
\hat{E}_{I}^{i}(x):\psi (A)\rightarrow -i\frac{\delta }{\delta A_{i}^{I}(x)}%
\psi (A).%
\end{array}
\label{Sch}
\end{equation}%
The commutators of these quantum operators mirror the classical Poisson
brackets, as required:%
\begin{eqnarray*}
\left[ \hat{A}_{i}^{I}(x),\hat{A}_{j}^{J}(y)\right] =0=\left[ \hat{E}%
_{I}^{i}(x),\hat{E}_{J}^{j}(y)\right] ,\left[ \hat{E}_{J}^{j}(y),\hat{A}%
_{i}^{I}(x)\right] =-i\delta ^{3}\left( x,y\right) \delta _{i}^{j}\delta
_{J}^{I}.
\end{eqnarray*}%
where we have set Planck's constant $\hbar $ equal to 1.

The motivation for choosing a canonical quantization lies in the hope of
addressing physical questions relating to ground states, and ultimately
measures on the space of field configurations, for quantized gauge theories.
\ In a canonical approach, gauge invariance is implemented at the quantum 
level by means of Dirac constraints. \ The ground state presented here is 
in fact automatically gauge-invariant (as well as spatially rotation and 
translation invariant) from its construction. \ Ideally, full Poincare 
invariance of the ground state is hoped to result by promoting all Poincare 
generators to quantum conserved quantities and verifying that they annihilate 
the ground state as well as exibiting appropriate commutators.
Formulated in terms of the Ashtekar variables, general relativity is also a 
gauge theory, and under a canonical quantization, full diffeomorphism invariance is similarly imposed as a quantum constraint operator (see e.g. \cite{Smolin}).

Yang-Mills theory is a valuable testing ground for ideas to be applied to canonical quantum general relativity. \ For instance the Kodama state, at present the only known candidate ground state for gravity in the Ashtekar variables (see \cite{Smolin}), arose as a generalization of the Chern-Simons state in Yang-Mills theory. \ While exhibiting many positive features, the Kodama state as usually constructed seems to inherit unphysical properties of the Chern-Simons state. \ Alternative candidate ground states for canonical quantum general relativity may aid the ongoing search for a physical inner product. \ Since the Kodama state is a generalization from the Chern-Simons ground state in Yang-Mills theory, a 
reasonable effort towards the construction of a normalizable ground state for 
quantum general relativity would be to search for a well-behaved ground state 
in Yang-Mills theory.

Such being the aim of the current project, we do not attempt to conform to the usual ideals of quantized Yang-Mills theory per se, as envisioned for instance in the formulation of the Clay Prize Millenium Problem \cite{JaffeWitten}. \ That is, we do not seek a quantization likely to yield a mass gap, since in quantum gravity a mass gap is not expected.

In the Schrodinger representation (\ref{Sch}), a ground state $\Omega \left(
A\right) $ for the Hamiltonian operator is the first step toward finding a
state space of the form $L^{2}\left( \mathcal{A},d\mu \right) ,$ for some
measure $d\mu $ on the space of connections $\mathcal{A}$. \ Heuristically
speaking, the first candidate for the measure $d\mu $ would be something like%
\begin{equation}
``\left[ \Omega \left( A\right) \right] ^{2}dA,"  
\label{YMmeasure}
\end{equation}%
where $``dA"$ is a naive ``Lebesgue'' measure on $\mathcal{A}$. \ Of course,
the usual way (e.g. \cite{GlimmJaffe}) to make sense of such expressions
will encounter resistance on two fronts: \ first, $\Omega \left( A\right) $
will be non-Gaussian for a nonabelian gauge theory, and secondly the space $%
\mathcal{A}$ should in fact be composed of equivalence classes of
connections modulo gauge transformations $\mathcal{A}/\mathcal{G}$, and
hence is not a linear space. \ Similar difficulties are at least partly
addressed within the literature on Yang-Mills path integrals (\cite%
{Rivasseau}, \cite{almmt}); however, arriving at a full rigorous
understanding of a measure such as (\ref{YMmeasure}) is obviously highly
nontrivial. \ In the meantime, however, it would be nice at least to know a
well-behaved candidate ground state for Yang-Mills theory. \ By
well-behaved, we mean that the ground state should decay rapidly for
connections which are ``large'' in some suitable sense -- e.g. of large $%
L^{2}$ norm -- so as to be a normalizable ground state with respect to some
measure on $\mathcal{A}$. \ This is what the Chern-Simons state%
\begin{eqnarray*}
\Psi _{CS}\left( A\right) =\exp \left[ \int_{\Sigma }tr\left( A\wedge dA-%
\frac{2}{3}A\wedge A\wedge A\right)d^{3}x\right] 
\end{eqnarray*}%
fails to do, since the Chern-Simons form in the exponent changes sign under
parity (for a good discussion of the Chern-Simons state's problems, see \cite%
{Witten}). \ For the abelian case of free Maxwell theory, a well-behaved
zero-energy ground state has already been written by Wheeler \cite%
{Geometrodynamics} in closed form as%
\begin{equation}
\Omega (A)= \mathcal{N}\exp \left( -\frac{1}{4\pi ^{2}}\int\limits_{%
\mathbb{R}^{3}}\int\limits_{\mathbb{R}^{3}}\frac{\left( \nabla \times
A(x)\right) \cdot \left( \nabla \times A(y)\right) }{\left| x-y\right| ^{2}}%
d^{3}x d^{3}y\right) ,  
\label{Wheeler}
\end{equation}%
and in fact for linearized general relativity, Kuchar \cite{Kuchar} derived
the strongly analogous ground state wave functional%
\begin{eqnarray*}
\Psi \left( h\right) =\mathcal{N}\exp \left( -\frac{1}{8\pi ^{2}}%
\int\limits_{\mathbb{R}^{3}}\int\limits_{\mathbb{R}^{3}}\frac{\left(
h_{ik,l}^{TT}(x)\right) \cdot \left( h_{ik,l}^{TT}(y)\right) }{\left|
x-y\right| ^{2}} d^{3}x d^{3}y\right) 
\end{eqnarray*}%
in terms of the linearized metric tensor%
\begin{eqnarray*}
h_{ik}=g_{ik}-\eta _{ik}
\end{eqnarray*}%
in transverse traceless gauge (denoted $h_{ik}^{TT}$).

The explicit construction of such ground states, however, relies on integral
kernel methods available only for linear theories. \ To deal with the
nonlinearities displayed by full nonabelian Yang-Mills theory or general
relativity, we need a new and more indirect means of finding well-behaved
ground states, reducing in free cases to these known examples.

Thus motivated, we generalize a method developed by Moncrief and Ryan (\cite%
{Moncrief}, \cite{MonRy}, \cite{Ryan}) in nonlinear quantum mechanical
settings, using classical Hamilton-Jacobi theory to derive an expression for
an exact quantum state which is a zero-energy ground state with respect to a
particular ordering of the Hamiltonian operator. \ Encouragement for the prospect of extending to general relativity comes from the fact that in \cite{MonRy}, Moncrief and Ryan present an explicit solution for such a ground state in the vacuum Bianchi IX cosmology.  

As explained in Sect.~\ref{Ordering},\ the ground state we seek is essentially the exponential of Hamilton's principal function for the corresponding Euclidean problem, so in order to find the principal function (or rather functional), we must solve the Dirichlet problem for Yang-Mills theory. \ For a compact base manifold, this has been collectively achieved by Uhlenbeck \cite{Uhlenbeck}, Sedlacek \cite{Sedlacek}, and Marini \cite{Marini}, using the direct method in the
calculus of variations. \ We follow a similar technique but generalize to
the case of a noncompact manifold, since we are interested in Yang-Mills
theory on Minkowski space. \ Some preliminaries necessary to solving the
Dirichlet problem on a Riemannian manifold are presented in Sect.~\ref{Prelim},
and the solution to the Riemannian Yang-Mills Dirichlet problem is presented
in Sect.~\ref{YMDirichlet}. \ Finally, in Sect.~\ref{Gauge}, we conclude gauge
invariance of the ground state and discuss partial results and ongoing work
to test Poincare invariance.

\section{Preliminaries}
\label{Prelim}

To solve the Yang-Mills Dirichlet problem for a compact manifold $M$, Marini %
\cite{Marini} introduces a terminology for coverings of $M$ by geodesic
balls and half-balls; these are described respectively as neighborhoods of
type 1 and type 2. \ Let $M$ be a smooth $n$-dimensional manifold $M$
equipped with a Riemannian metric $g$, and let $\partial M$ be its boundary.
\ Then neighborhoods of type 1, in the manifold's interior, are denoted%
\begin{eqnarray*}
U^{\left( 1\right) }\equiv \left\{ x=\left( x^{0},...,x^{n-1}\right) :\left|
x\right| <1\right\} 
\end{eqnarray*}%
while neighborhoods of type 2, centered around points in $\partial M$, are
of the form%
\begin{eqnarray*}
U^{\left( 2\right) }\equiv \left\{ x=\left( x^{0},...,x^{n-1}\right) :\left|
x\right| <1,x^{0}\geq 0\right\} ,
\end{eqnarray*}%
where the coordinate $x^{0}$ parametrizes unit-speed geodesics orthogonal to 
$\partial M=\left\{ x^{0}=0\right\} $. \ The boundary of a type 2
neighborhood divides into 
\begin{eqnarray*}
\partial _{1}U &=&\left\{ x\in \partial U^{\left( 2\right) }:x^{0}=0\right\}
, \\
\partial _{2}U &=&\left\{ x\in \partial U^{\left( 2\right) }:\left| x\right|
=1\right\} .
\end{eqnarray*}

In our problem, the manifold of interest is $\mathbb{R}_{+}\times 
\mathbb{R}^{3}=\left\{ \left( x^{0},x^{1},x^{2},x^{3}\right) :x^{0}\geq
0\right\} $ with the Euclidean metric; however we solve the Yang-Mills
Dirichlet problem for a general smooth 4-dimensional Riemannian manifold
with boundary, generalizing Marini's procedure to the non-compact case (Sect.~\ref{YMDirichlet}). \ Certain results used are also valid in general
dimension $n$; such distinctions are clearly noted in the statements. \ We
return to consider the importance of dimension more thoroughly in Sect.~\ref%
{YMDirichlet}.

The main ingredient in a Yang-Mills theory is the structure group; this is a
compact Lie group $G \subset SO \left( l \right)$ with Lie algebra $\mathfrak{g}$. \ For $P$ a principal 
$G$-bundle over $M$, the Yang-Mills field is a connection $A\in \Lambda
^{1}P\otimes \mathfrak{g}$. \ Given a local section $\sigma _{\alpha
}:U_{\alpha }\rightarrow P$ for some neighborhood $U_{\alpha }\subset M$,
the connection 1-form $A$ pulls back to a $\mathfrak{g}$-valued 1-form $%
A_{\alpha }=\sigma _{\alpha }^{\ast }A$ on $U_{\alpha }$; the transformation
of $A$ on overlapping neighborhoods $U_{\alpha }$ and $U_{\beta }$ is given
by the transition function $\tau _{\alpha \beta }:U_{\alpha }\cap U_{\beta
}\rightarrow G$ defined by $\sigma _{\beta }\left( x\right) =\sigma _{\alpha
}\left( x\right) \tau _{\alpha \beta }(x)$:%
\begin{eqnarray*}
A_{\alpha }(x)=\tau _{\alpha \beta }(x)^{-1}d\tau _{\alpha \beta }(x)+\tau
_{\alpha \beta }(x)^{-1}A_{\beta }\left( x\right) \tau _{\alpha \beta }(x),\    
\ x\in U_{\alpha }\cap U_{\beta }.
\end{eqnarray*}%
The important quantity for Yang-Mills theory is the curvature $F\in \Lambda
^{2}P\otimes \mathfrak{g}$ of the connection $A$, given by $F=d_{P}A+\frac{1%
}{2}\left[ A,A\right] $ where the bracket $\left[ \cdot ,\cdot \right] $
denotes the graded commutator on forms, so that $\left[ A,A\right] =2\left(
A\wedge A\right) .$ \ In terms of a local section $\sigma _{\alpha
}:U_{\alpha }\rightarrow P$, $F$ pulls back to a $\mathfrak{g}$-valued
2-form $F_{\alpha }=\sigma _{\alpha }^{\ast }F$ on $U_{\alpha }$, given by $%
F_{\alpha }=d_{M}A_{\alpha }+\frac{1}{2}\left[ A_{\alpha },A_{\alpha }\right]
$, transforming as%
\begin{eqnarray*}
F_{\alpha }(x)=\tau _{\alpha \beta }(x)^{-1}F_{\beta }\left( x\right) \tau
_{\alpha \beta }\left( x\right) ,\    \ x\in U_{\alpha }\cap U_{\beta }
\end{eqnarray*}%
for $\tau _{\alpha \beta }$ as given above. \ In local coordinates, $F$
reduces to%
\begin{eqnarray*}
F_{\mu \nu }=\partial _{\mu }A_{\nu }-\partial _{\nu }A_{\mu }+\left[ A_{\mu
},A_{\nu }\right] .
\end{eqnarray*}%
(here $\left[ \cdot ,\cdot \right] $ is the ordinary commutator in $%
\mathfrak{g}$).%

In order to describe the Yang-Mills action, we use the local expressions of $%
F$ as a $\mathfrak{g}$-valued 2-form on neighborhoods of $M$; however all
definitions are gauge-invariant and therefore do not depend on the
particular section used to pull back $F$.

The Yang-Mills action can be conveniently couched in terms of the inner
product for $\mathfrak{g}$-valued $k$-forms on the manifold $M$:%
\begin{equation}
\left\langle \eta ,\theta \right\rangle _{2}=\int_{M}tr\left( \eta \wedge
\ast \theta \right) ,  
\label{FormProd}
\end{equation}%
where $\ast $ denotes the Hodge dual with respect to the metric $g$ on $M$.
\ We occasionally also write $\left\langle \eta ,\theta \right\rangle $ for
the pointwise inner product $\left\langle \eta ,\theta \right\rangle
=tr\left( \eta \wedge \ast \theta \right) $. \ The inner product (\ref%
{FormProd}) in turn allows us to define $L^{p}$ and Sobolev spaces of forms,
using the norm%
\begin{eqnarray*}
\left\| \eta \right\| _{p}=\left( \int_{M}\left| \eta \right|
^{p}\right) ^{\frac{1}{p}}=\left( \int_{M}\left\langle \eta ,\eta \right\rangle
^{p/2}\right) ^{\frac{1}{p}}=\left( \int_{M}\left[ tr\left( \eta \wedge
\ast \eta \right) \right] ^{p/2}\right) ^{\frac{1}{p}}.
\end{eqnarray*}%
In terms of local coordinates $\left\{ x^{\mu} \right\}$ on $M$ and a basis $\left\{ e_{I} \right\}$ for the Lie algebra $\mathfrak{g}$, membership of $\eta$ in the Sobolev space of forms is equivalent to each component $\eta^{I}_{\mu_1 ... \mu_k}$ being Sobolev in the ordinary sense of functions. \ Using this notation, the Yang-Mills action can be given as%
\begin{eqnarray*}
I(A)=\frac{1}{2}\left\| F\right\| _{2}^{2}=\frac{1}{2}\int_{M}tr\left(
F\wedge \ast F\right) =\frac{1}{4}\int_{M}trF_{\mu \nu }F^{\mu \nu }\sqrt{g}%
dx^{1}\cdot \cdot \cdot dx^{n},
\end{eqnarray*}%
where integration is done using the local form of $F$ on a neighborhood
(gauge invariance of the form $tr\left( F\wedge \ast F\right) $ negates any
ambiguity due to choice of local trivialization).

The above manner of formulating the Yang-Mills action functional also offers
an easy proof of lower semicontinuity, necessary in using the direct method
to find a minimizer:

\begin{theorem}
The Yang-Mills functional on a manifold $M$ of dimension 4 is lower
semicontinuous with respect to the weak topology on $W_{loc}^{1,2}\left(
M\right) .$
\end{theorem}

\begin{proof}
It is good enough to prove that on any open bounded set $U\subset M$, if $%
A_{i}\rightharpoonup A$ in $W^{1,2}\left( U\right) $, then $I\left( A\right)
\leq \lim \inf_{i\rightarrow \infty }I\left( A_{i}\right) $. \ Locally we
can write%
\begin{eqnarray*}
F_{i}=dA_{i}+\frac{1}{2}\left[ A_{i},A_{i}\right] .
\end{eqnarray*}%
Using the same reasoning as Sedlacek's in Lemma 3.6 of \cite{Sedlacek}, weak
convergence of $\left\{ A_{i}\right\} $ to $A$ in $W^{1,2}\left( U\right) $
implies weak convergence of $dA_{i}$ to $dA$ in $L^{2}\left( U\right) .$ \
The continuity of the imbedding $W^{1,2}\hookrightarrow L^{4}$ and of the
multiplication $L^{4}\times L^{4}\rightarrow L^{2}$ along with boundedness
of $\left\{ \left| \left| A_{i}\right| \right| _{2,1} \right\} $ implies that 
$\left\{ \left| \left| \left[ A_{i},A_{i}\right] \right| \right|
_{2}\right\} $ is bounded. \ This together with a.e. pointwise convergence
yield $\left[ A_{i},A_{i}\right] \rightharpoonup \left[ A,A\right] $, so
that $F_{i}\rightharpoonup F$ in $L^{2}\left( U;P\right) $. \ Finally, lower
semicontinuity of the $\left| \left| \cdot \right| \right| _{2}$ norm
concludes lower semicontinuity of the Yang-Mills functional.
\end{proof}

The Yang-Mills field equations are%
\begin{eqnarray*}
d_{D}\ast F=0
\end{eqnarray*}%
where $d_{D}=d+\left[ A,\cdot \right] $; solutions to this system correspond
exactly to critical points of the Yang-Mills action $I(A)$. \ To see this,
vary $I(A)$ by varying $A$ as $A+\lambda h$, where $h$ vanishes at $t=0$ and
is supported on some compact subset $N$ (dependent on $h$) of $M$:%
\begin{eqnarray*}
\delta _{h}\left( I\right) \left( A\right)  &=&\int_{N}\left\langle
d_{D}h,F\right\rangle =\int_{N}tr\left( d_{D}h\wedge \ast F\right)  \\
&=&\int_{\partial N}tr\left( h\wedge \ast F\right) -\int_{N}tr\left( h\wedge
d_{D}\ast F\right) .
\end{eqnarray*}%
It is evident that $\delta _{h}\left( I\right) \left( A\right) $ vanishes
for all variations $h$ precisely when $d_{D}\ast F$ is identically 0.

As described in Sect.~\ref{Intro}, this paper will deal with the canonical
quantization ansatz; therefore we must derive the canonical variables. \ To
make the transformation from a Lagrangian to a Hamiltonian formulation, we
specialize to the case of interest $M=\mathbb{R}_{+}\times \mathbb{R}%
^{3}$. \ Since $M$ is contractible, every bundle over $M$ is trivial and
therefore admits a global section. \ We can then drop the distinction
between $A$ and its local coordinate representation. \ Because the
Lagrangian is independent of $A_{0}$, the Legendre transformation breaks
down for an arbitrary gauge (see e.g. \cite{Jackiw}), and we must choose to
work in the Weyl gauge $A_{0}=0.$ \ Thus our canonical position variable is $%
A_{i}^{I}$, where $i$ runs over the three spatial parameters and $I$ over
the basis of the Lie algebra $\mathfrak{g}$, and the canonical momentum is $%
E_{I}^{i}=\dot{A}_{i}^{I}$ (this is the negative of the ``electric field''
variable).

With respect to these variables we obtain the Hamiltonian%
\begin{equation}
H=\frac{1}{2}\int_{\mathbb{R}^{3}}tr\left( E^{2}+B^{2}\right) \ ,
\label{YMHam}
\end{equation}%
where $B^{i}=\frac{1}{2}\varepsilon ^{ijk}F_{jk}$. \ Hamilton's equations
follow by writing the integral $J=\int_{0}^{\infty }H$ $dt$ as $%
\int_{0}^{\infty }\int_{\mathbb{R}^{3}}\mathcal{H}$ \ $d^{3}x$ $%
dt=-I+\int_{0}^{\infty }\int_{\mathbb{R}^{3}}E\dot{A}$ \ $d^{3}x$ $dt$. \
Varying both expressions with respect to a one-parameter family $A_{\lambda }
$ (where the variation has compact support in $\mathbb{R}_{+}\times 
\mathbb{R}^{3}$ and vanishes for $t=0$), we arrive at the equality%
\begin{eqnarray*}
\frac{dJ}{d\lambda } &=&\int_{0}^{\infty }\int_{\mathbb{R}^{3}}\frac{\delta H%
}{\delta A}\delta A+\frac{\delta H}{\delta E}\delta E \\
&=&\int_{0}^{\infty }\int_{\mathbb{R}^{3}}\left[ E\delta \dot{A}+\dot{A}%
\delta E\right] -\frac{dI}{d\lambda } \\
&=&\int_{0}^{\infty }\int_{\mathbb{R}^{3}}\left[ -\dot{E}\delta A+\dot{A}%
\delta E\right] -\frac{dI}{d\lambda },
\end{eqnarray*}%
using integration by parts. \ In order for equality to hold between the
first and last lines for all variations, Hamilton's equations%
\begin{eqnarray*}
\dot{E} &=&-\frac{\delta H}{\delta A} \\
\dot{A} &=&\frac{\delta H}{\delta E}
\end{eqnarray*}%
must be equivalent to the vanishing of the variation $\frac{\delta I}{\delta
A}.$
Notice that in taking the Weyl gauge $A_0 = 0$, we have lost the field equations describing gauge transformations, and therefore in the quantized theory, the Gauss law constraint
\begin{eqnarray*}
D_{i}E^{i} = 0
\end{eqnarray*}%
must be dealt with separately, either by promoting to a quantum operator and verifying that it annihilates the ground state, or by other means.  For the ground state we construct here, gauge invariance in fact turns out the be directly verifiable (see Sect.~\ref{Gauge}).

\section{Nonlinear normal ordering}
\label{Ordering}

For nonlinear quantum mechanical situations, Moncrief \cite{Moncrief} and
Ryan \cite{Ryan} present a ``normal'' ordering scheme for the Hamiltonian
operator, yielding a well-behaved associated ground state. \ Consider a
nonlinear quantum mechanical Hamiltonian of the form%
\begin{eqnarray*}
H=\frac{1}{2}\left| p\right| ^{2}+V(x).
\end{eqnarray*}%
In a linear system, the function $V(x)$ would be a quadratic form $%
\left\langle x,Mx\right\rangle $, $M$ a positive self-adjoint operator. \
The normal ordering would proceed by factoring $M$ as $T^{2}$, in terms of
its unique positive self-adjoint square root $T$, and defining creation and
annihilation operators $a^{\ast }=\frac{1}{\sqrt{2}}\left( T\hat{x}-i\hat{p}%
\right) $ and $a=\frac{1}{\sqrt{2}}\left( T\hat{x}+i\hat{p}\right) $. \
Under usual assignment of canonical quantum operators $\hat{x}^{i}:\psi
(x)\rightarrow x^{i}\psi (x)$, $\hat{p}^{i}:\psi (x)\rightarrow -i\frac{%
\partial }{\partial x^{i}}\psi (x)$, this immediately yields the ground
state $\psi (x)=\mathcal{N}\exp \left( -\frac{\left\langle x\ ,%
Tx\right\rangle }{2}\right) $ with energy $\frac{tr\ T}{2}$, for $H$
expressed as a quantum operator $\hat{H}=a^{\ast }a+\frac{tr\ T}{2}I.$

The idea of the nonlinear normal ordering is to factorize the function $V(x) \geq 0$ by solving the imaginary-time zero-energy Hamilton-Jacobi equation%
\begin{equation}
\frac{1}{2}\sum\limits_{i}\left( \frac{\partial S}{\partial x^{i}}\right)
^{2}-V(x)=0.  
\label{HJE}
\end{equation}%
We can then order the quantum Hamiltonian operator as%
\begin{equation}
\hat{H}=\frac{1}{2}\sum\limits_{i}\left( \widehat{\frac{\partial S}{\partial
x^{i}}}-i\hat{p}^{i}\right) \left( \widehat{\frac{\partial S}{\partial x^{i}}%
}+i\hat{p}^{i}\right) ,  
\label{NNO}
\end{equation}%
admitting the zero-energy ground state%
\begin{eqnarray*}
\mathcal{N}\exp \left( -S(x)\right) .
\end{eqnarray*}%
This factorization can be illustrated with the anharmonic oscillator
Hamiltonian%
\begin{equation}
H=\frac{1}{2}p^{2}+\frac{1}{2}x^{2}+\frac{1}{4}\lambda x^{4}\ ,
\label{NO}
\end{equation}%
in which case the Hamilton-Jacobi equation $\frac{1}{2}\left( \frac{dS}{dx}%
\right) ^{2}=\frac{1}{2}x^{2}+\frac{1}{4}\lambda x^{4}\ $is easily
integrated to yield the solution$\ S(x)=\frac{2}{3\lambda }\left( 1+\frac{%
\lambda }{2}x^{2}\right) ^{3/2}-\frac{2}{3\lambda }.$ \ While the resulting
ground state is not the usual anharmonic oscillator ground state obtained
from the factor ordering $\hat{H}=\frac{1}{2m}\hat{p}^{2}+\frac{m\omega ^{2}%
}{2}\hat{q}^{2}+\frac{1}{4}\lambda \hat{q}^{4}$, it is the correct
zero-energy ground state for nonlinear normal ordering (\ref{NNO}). \
Finding a ground state with zero energy is not a priority for an ordinary
quantum mechanical system like the anharmonic oscillator, but in the realm
of relativistic field theories, a quantum ground state must have zero energy
to be invariant under infinitesimal time-translations. \ Hence full Poincare
invariance in a canonically quantized relativistic field theory requires a
zero energy ground state, suggesting that the nonlinear normal ordering
approach may yield well-behaved candidate ground states for a canonically
quantized nonlinear field theory.

Following this line of reasoning for Yang-Mills theory, we set up the
imaginary-time zero-energy Hamilton-Jacobi equation for the Yang-Mills
Hamiltonian (\ref{YMHam}):
\begin{equation}
\int_{\mathbb{R}^{3}}tr\left( \frac{\delta S}{\delta A}\right) ^{2}=\int_{%
\mathbb{R}^{3}}trB^{2},  
\label{YMHJE}
\end{equation}%
whose solution $S\left( A\right) $ will yield the nonlinear normal ordering%
\begin{eqnarray*}
\hat{H}=\frac{1}{2}\int_{\mathbb{R}^{3}}tr\left( \widehat{\frac{\delta S}{%
\delta A}}-i\hat{E}\right) \left( \widehat{\frac{\delta S}{\delta A}}+i%
\hat{E}\right)
\end{eqnarray*}%
and associated zero-energy ground state%
\begin{eqnarray*}
\mathcal{N}\exp \left( -S(A)\right) .
\end{eqnarray*}%
However, the Hamilton-Jacobi equation (\ref{YMHJE}) is not as easily solved
as in the anharmonic oscillator problem. \ Fortunately, the classical
Hamilton-Jacobi theory provides us with a means of constructing the
solution, essentially as Hamilton's principal function for the
imaginary-time problem. \ With the transformation to imaginary time $%
t\rightarrow it$, the chain rule yields $A\rightarrow A$, $\dot{A}%
\rightarrow -i\dot{A}$, and $E\rightarrow -iE$, so that the imaginary-time
Lagrangian and Hamiltonian are%
\begin{eqnarray*}
\tilde{L} &=&\frac{1}{2}\int_{\mathbb{R}^{3}}\left( -\dot{A}%
^{2}-B^{2}\right) \ d^{3}x \\
\tilde{H} &=&\frac{1}{2}\int_{\mathbb{R}^{3}}\left( -E^{2}+B^{2}\right) \ d^{3}x.
\end{eqnarray*}%
The full Hamilton-Jacobi equation is%
\begin{equation}
\frac{\partial S}{\partial t}+\tilde{H}\left( A,\frac{\delta S}{\delta A}%
,t\right) =0;  
\label{tHJE}
\end{equation}%
a time-independent solution $S(A)$ to this equation will be the solution we
seek for (\ref{YMHJE}). \ In fact, the solution we construct will be
time-independent, but for the moment we assume that an explicit time
dependence is possible, using the definition of functional derivatives to
write%
\begin{eqnarray*}
\frac{dS}{dt}=\frac{\partial S}{\partial t}+\int_{\mathbb{R}^{3}}\frac{%
\delta S}{\delta A_{t}}\frac{\partial A_{t}}{\partial t}\ d^{3}x.
\end{eqnarray*}%
Subsituting from (\ref{tHJE}) we obtain%
\begin{eqnarray*}
\frac{dS}{dt} &=&-\tilde{H}\left( A_{t},\frac{\delta S}{\delta A_{t}}\right)
+\int_{\mathbb{R}^{3}}\frac{\delta S}{\delta A_{t}}\frac{\partial A_{t}}{%
\partial t}\ d^{3}x \\
&=&\int_{\mathbb{R}^{3}}\frac{\delta S}{\delta A_{t}}\frac{\partial A_{t}}{%
\partial t}-\tilde{\mathcal{H}}\left( A_{t},\frac{\delta S}{\delta A_{t}}%
\right) \ d^{3}x.
\end{eqnarray*}%
For $A_{t}$ the solution to%
\begin{equation}
\frac{\partial A_{t}}{\partial t}=\left. \frac{\delta \tilde{H}}{\delta E}%
\medskip \right| _{E=\frac{\delta S}{\delta A_{t}}}
\end{equation}%
with initial data $A_{t=0}=A$, we get%
\begin{eqnarray*}
\int_{\mathbb{R}^{3}}\frac{\delta S}{\delta A_{t}}\frac{\partial A_{t}}{%
\partial t}-\tilde{\mathcal{H}}\left( A_{t},\frac{\delta S}{\delta A_{t}}%
\right) \ d^{3}x &=&\int_{\mathbb{R}^{3}}E\dot{A}_{t}-\tilde{\mathcal{H}}%
\left( A_{t},E\right) \ d^{3}x \\
&=&L\left( A_{t},DA_{t}\right) \\
&\Rightarrow &S\left( A_{t_{0}}\right) -S\left( A\right) =\int_{0}^{t_{0}}%
\tilde{L}\left( A_{t},\dot{A}_{t}\right) \ dt
\end{eqnarray*}

Taking $S\left( A\right) =-\int_{0}^{\infty }\tilde{L}\left( A_{t},\dot{A}%
_{t}\right) $ $dt$ clearly satisfies this relation, since we will then have $%
S\left( A_{t_{0}}\right) =-\int_{t_{0}}^{\infty }\tilde{L}\left( A_{t},\dot{A%
}_{t}\right) $ $dt$. \ The exponential $\exp \left( -S(A)\right) $ will peak
about the field configuration $A=0$. \ To prove that the functional $S(A)$
exists, we need only prove the existence of a solution $A_{t}$ of the
Euclidean Yang-Mills equations, with initial data $A$.

\section{Solving the Yang-Mills Dirichlet problem}
\label{YMDirichlet}

For a compact manifold $M$ with boundary $\partial M$, the method of
Uhlenbeck \cite{Uhlenbeck}, Sedlacek \cite{Sedlacek}, and Marini \cite%
{Marini} begins with a localizing theorem, proving that given a sequence of
connections with a uniform global bound on the Yang-Mills action, there
exists a cover for $M$ (possibly missing a finite collection of points) such
that on neighborhoods of the cover, the Yang-Mills action for connections in
the sequence eventually becomes lower than an arbitrary pre-set bound $%
\varepsilon $. \ This result depends on compactness as proved in \cite%
{Sedlacek} (see Proposition 3.3, or in \cite{Marini} Theorem 3.1). \ We
reprove the result here in a manner independent of compactness (Theorem \ref%
{goodcover}), so that the overall argument now applies to noncompact
manifolds with boundary as well.

Note that another possible solution to the problem would be to transform $M=%
\mathbb{R}_{+}\times \mathbb{R}^{3}$ into a compact manifold with
boundary, using inversion in the sphere (this suggestion is due to T.
Damour). \ In this approach, one considers the unit sphere centered at the
origin of $\mathbb{R}^{4}$, imbedding $\mathbb{R}_{+}\times \mathbb{R}%
^{3}$ into $\mathbb{R}^{4}$ as the set $\left\{ x:x^{0}\geq 2\right\} $. \
The inversion mapping is%
\begin{eqnarray*}
y^{i}=\frac{x^{i}}{r^{2}}\ ,
\end{eqnarray*}%
where $r^{2}=\left( x^{0}\right) ^{2}+\left( x^{1}\right) ^{2}+\left(
x^{2}\right) ^{2}+\left( x^{3}\right) ^{2}$. \ Under this transformation,
the hyperplane $x^{0}=2$ maps to a sphere $S_{1/4}$ of radius $\frac{1}{4}$,
with its south pole (the image of all points at infinity) at the origin. \
The half-space $\left\{ x:x^{0}>2\right\} $ maps to the interior of $S_{1/4}$.

Since the mapping is conformal, the Yang-Mills action remains invariant, and
the problem of interest on $\mathbb{R}_{+}\times \mathbb{R}^{3}$ has
been effectively mapped to a compact problem, to which the arguments of \cite%
{Uhlenbeck}, \cite{Sedlacek}, and \cite{Marini} should apply directly. \ We
do not pursue this approach here, since the crucial result using compactness
can be shown to generalize (Theorem \ref{goodcover}); however we note its
potential usefulness to future work, such as the investigation of uniqueness
of solution to the Yang-Mills Dirichlet problem (see Sect.~\ref{Fun}). \ An
issue pertinent to the conformal mapping approach is the behavior of initial
data at the south pole of $S_{1/4}$, the image of points at infinity.

Returning to our sketch of the Yang-Mills Dirichlet problem's solution,
local control over the Yang-Mills action is used to prove existence and
regularity of a minimizer. \ From this point on, the proofs in \cite%
{Uhlenbeck}, \cite{Sedlacek}, and \cite{Marini} are purely local and hold
unchanged in the noncompact case; proofs are thus not repeated here. \
Locally, the argument for existence of a Yang-Mills minimizer consists in\
finding a Sobolev-bounded minimizing sequence satisfying the boundary
conditions; this sequence then has a weakly convergent subsequence, which
proves to be a solution to the original Dirichlet problem. \ Local solutions
are related by transition functions on overlapping neighborhoods.

Gauge freedom turns out to be a help as well as a hindrance. \ Of course it
forces the necessity of working locally and proving compatibility on
overlaps, but at the same time gauge freedom offers an elegant solution to
the regularity problem. \ A judicious choice of gauge -- the ``Hodge gauge''
-- complements the Yang-Mills equation in such a way as to yield an elliptic
system. \ In the Yang-Mills equation%
\begin{eqnarray*}
d\ast dA+\left[ A,dA\right] +\left[ A,\left[ A,A\right] \right] ,
\end{eqnarray*}%
the highest order term is related to the first term of the Laplace-de Rham
operator $\Delta =\delta d+d\delta $, where $\delta $ is the codifferential $%
\delta =\left( -1\right) ^{n(k+1)+1}\ast d\ast $ ($k$ being the degree of
the differential form operated upon). \ Choosing the Hodge gauge, in which $%
d\ast A=0$, ensures that every solution of our system in this gauge is also
a solution of the elliptic system $\Delta A+\ast \left( \left[ A,dA\right] +%
\left[ A,\left[ A,A\right] \right] \right) =0$, and therefore enjoys the
regularity properties of such solutions. \ (Additional work is needed to
establish boundary regularity; Marini accomplishes this in \cite{Marini}
using the technique of local doubling.)

In the physical problem, we are interested in Yang-Mills theory over a $4$%
-dimensional manifold with boundary. \ However many theorems which follow
are also valid over any smooth $n$-dimensional Riemannian manifold with
boundary, and we retain this level of generality in stating and proving
results. \ The caveat lies in stringing together the individual theorems
into a complete argument for existence and regularity of a solution to the
Yang-Mills Dirichlet problem; to accomplish this, the dimension must be 4
(see the remarks in \cite{Marini} following Theorem 3.1). \ The ``good
cover'' theorem (Theorem \ref{goodcover} here, or Theorem 3.1 in \cite{Marini}%
) guarantees a cover of $M\backslash \left\{ x_{1},...,x_{k}\right\} $ on
whose neighborhoods the local Yang-Mills action for the connections in the
sequence is eventually bounded by an arbitrary pre-set bound $\varepsilon $.
\ However, the condition for existence of a Hodge gauge solution to the
Yang-Mills Dirichlet problem is a bound on the local $L^{n/2}$ norm of the
Yang-Mills field strength $F$, which except in dimension $4$ is not the same
as a local bound on the Yang-Mills action.

Without further ado, we give the precise statements of all theorems needed
for the existence and regularity of a Yang-Mills minimizer on a $4$%
-dimensional manifold with boundary. \ On a neighborhood $U$ of type 1 or 2,
the condition for local existence of a gauge satisfying $d\ast A=0$ is an $%
L^{n/2}$ bound on Yang-Mills field strength. \ Consider the sets%
\begin{eqnarray*}
\mathfrak{A}_{K}^{1,p}(U) &=&\left\{ D=d+A:\   \ A\in W^{1,p}(U), \   \ \left\|F_{A}\right\| _{L^{n/2}(U)}<K\right\}  \\
\mathfrak{B}_{K}^{1,p}(U) &=&\left\{ D=d+A:\ \ 
\begin{array}{l}
A\in W^{1,p}(U) \\ 
A_{\tau }\in W^{1,p}(\partial _{1}U)%
\end{array}%
\ ,\  \ 
\begin{array}{l}
\left\| F_{A}\right\| _{n/2}<K \\ 
\left\| F_{A_{\tau }}\right\| _{L^{n/2}(\partial _{1}U)}<K%
\end{array}%
\right\} 
\end{eqnarray*}%
describing connections with field strength locally $L^{n/2}$-bounded on a
neighborhood $U$ of type 1 and type 2, respectively. \ (All norms are
defined on $U$, unless otherwise specified.) \ As proven in \cite{Uhlenbeck}
(Thm 2.1) for interior neighborhoods and in \cite{Marini} (Thms 3.2 and 3.3)
for boundary neighborhoods, a good choice of gauge exists for connections
belonging to $\mathfrak{A}_{K}^{1,p}(U)$ or $\mathfrak{B}_{K}^{1,p}(U).$ \
More precisely,

\begin{theorem}
\label{localgauge}For $\frac{n}{2}\leq p<n$, there exists $K\equiv K(n)>0$
and $c\equiv c(n)$ such that every connection $D=d+A\in \mathfrak{A}_{K}^{1,p}\left( U\right) $ ($\mathfrak{B}_{K}^{1,p}\left( U\right) $) is
gauge equivalent to a connection $d+\hat{A},$ $\hat{A}\in W^{1,p}(U),$ satisfying 
\\
$%
\begin{array}{l}
\begin{array}{ll}
(a)\  \ d\ast \hat{A}=0 \     \ &  (a^{\prime })\  \ \left(
\begin{array}{l}
d\ast \hat{A}=0 \\ 
d_{\tau }\ast \hat{A}_{\tau }=0\ \ on\ \partial _{1}U%
\end{array}%
\right)  \\ 
(b)\  \ \hat{A}_{\nu }=0 \ on \ \partial U \     \ & (b^{\prime })\  \ \hat{A}_{\nu }=0 \ on \ \partial _{2}U%
\end{array}
\\ 
(c=c^{\prime }) \  \ \left\| \hat{A}\right\| _{1,n/2}<c(n)\left\| F_{\hat{A}}\right\| _{n/2} \\ 
(d=d^{\prime }) \  \ \left\| \hat{A}\right\| _{1,p}<c(n)\left\| F_{\hat{A}}\right\| _{p}%
\end{array}%
$ \\
(Unprimed conditions $(a)$-$(d)$ refer to $\mathfrak{A}_{K}^{1,p}(U)$;
primed conditions $(a^{\prime })$-$(d^{\prime })$ to $\mathfrak{B}_{K}^{1,p}(U)$). \ Moreover, the gauge transformation $s$ satisfying $\hat{A}=s^{-1}ds+s^{-1}As$ can be taken in $W^{2,n/2}(U)$ ($s$ will in fact always be one degree smoother than $A$; see Lemma 1.2 in \cite{Uhlenbeck}).
\end{theorem}%

\begin{proof}
See \cite{Uhlenbeck}, \cite{Marini}.
\end{proof}

As noted in \cite{Marini}, the condition $\left\| F_{A}\right\| _{n/2}<K$ is
conformally invariant, while the norm $\left\| F_{A_{\tau }}\right\|
_{L^{n/2}(\partial _{1}U)}$ picks up a factor of $r$ under the dilation $%
x^{\prime }=rx$, so that the simultaneous conditions $\left\| F_{A}\right\|
_{n/2}<K$, $\left\| F_{A_{\tau }}\right\| _{L^{n/2}(\partial _{1}U)}<K$ on a
neighborhood $U$ of type 2 can always be achieved by applying an appropriate
dilation (the Dirichlet boundary data is prescribed to be smooth, so $%
\left\| F_{A_{\tau }}\right\| _{L^{n/2}(\partial _{1}U)}$ already satisfies
some bound).

To find a regular minimizer of the Yang-Mills action on a 4-dimensional
manifold $M$, we must find a cover $\left\{ U_{\alpha }\right\} $ of $M$ and
a minimizing sequence $\left\{ A_{i}\right\} $ whose members satisfy%
\begin{eqnarray*}
S_{YM}(\left. A_{i}\right| _{U_{\alpha }})=\int_{U_{\alpha }}\left|
F_{A_{i}}\right| ^{2} dx<K\  \ \forall \ \alpha ,i,
\end{eqnarray*}%
where $K\equiv K(4)$ is as given in Theorem \ref{localgauge}. \ For a compact
manifold this is proved in \cite{Sedlacek} using a counting argument. \ Here
we use dilations of the neighborhoods in a cover to construct a proof valid
for any smooth Riemannian manifold.

\begin{theorem}
\label{goodcover}
Let $\left\{ A(i)\right\} $ be a sequence of connections in 
$G$-bundles $P_{i}$ over $M$, with uniformly bounded action $\int_{M}\left|
F(i)\right| ^{2}$ $dx<B$ \ $\forall $ $i$. \ For any $\varepsilon >0$, there
exists a countable collection $\left\{ U_{\alpha }\right\} $ of
neighborhoods of type 1 and 2, a collection of indices $I_{\alpha }$, a
subsequence $\left\{ A(i)\right\} _{\mathcal{I}^{\prime }}\subset \left\{
A(i)\right\} _{\mathcal{I}}$, and at most a finite number of points $\left\{
x_{1},...,x_{k}\right\} \in M$ such that%
\begin{eqnarray*}
\bigcup U_{\alpha } &\supset &\left. M\right\backslash \left\{
x_{1},...,x_{k}\right\} \\
\int_{U_{\alpha }}\left| F(i)\right| ^{2} dx &<&\varepsilon \  \ \forall i\in \mathcal{I}^{\prime }, \  \ i>I_{\alpha }.
\end{eqnarray*}
\end{theorem}

\begin{proof}
For each $n\in \mathbb{N},$ consider the cover $\left\{ B_{n}(x):x\in
M\right\} $ of $M$ given by geodesic balls of radius $\frac{1}{n}$ centered
at each point $x\in M$ (for $x\in \partial M$, the geodesic ``ball'' $%
B_{n}(x)$ will actually be a half-ball, a fact which makes no difference in
the proof). \ By separability, each such cover has a countable subcover $%
C_{n}=\left\{ B_{n}(x_{n,m}):m\in \mathbb{N}\right\} $.

On any ball $B_{n}(x_{n,m})$, we have the uniform bound $%
\int_{B_{n}(x_{n,m})}\left| F(i)\right| ^{2}$ $dx<B$ \ $\forall $ $i$. \
Therefore for the ball $B_{n}(x_{n,1})$ in a given cover $C_{n}$, there
exists a subsequence of $\left\{ A(i)\right\} $ for which the corresponding
subsequence of $\left\{ \int_{B_{n}(x_{n,1})}\left| F(i)\right|
^{2}dx\right\} $ converges. \ Of this subsequence, there exists a further
subsequence such that the corresponding subsequence of $\left\{
\int_{B_{n}(x_{n,2})}\left| F(i)\right| ^{2}dx\right\} $ converges, and so
on, for every $m$. \ Diagonalizing\footnote{%
Diagonalizing over a list of sequences $\left\{ a_{j}\left( i\right)
\right\} $ such as $%
\begin{array}{llll}
a_{1}\left( 1\right)  & a_{1}\left( 2\right)  & a_{1}\left( 3\right)  & 
\ldots  \\ 
a_{2}\left( 1\right)  & a_{2}\left( 2\right)  & a_{2}\left( 3\right)  & 
\ldots  \\ 
a_{3}\left( 1\right)  & a_{3}\left( 2\right)  & a_{3}\left( 3\right)  & 
\ldots  \\ 
\vdots  & \vdots  & \vdots  & \ddots 
\end{array}%
$ selects out the new sequence $\left\{ a_{i}\left( i\right) \right\} $. \
In the case important for this proof, each row represents a subsequence of
the previous row, so that for any $j$, the diagonalized sequence $\left\{
a_{k}\left( k\right) \right\} $is a subsequence of $\left\{ a_{j}\left(
i\right) \right\} $ for $k\geq j$.} over these nested subsequences, we
obtain a subsequence of $\left\{ A(i)\right\} $ such that the corresponding
subsequence of $\left\{ \int_{B_{n}(x_{n,m})}\left| F(i)\right|
^{2}dx\right\} $ converges for every $m\in \mathbb{N}$.

Performing a similar diagonalization over all covers $C_{n}$, there exists a
subsequence $\left\{ A(i)\right\} _{\mathcal{I}^{\prime }}\subset \left\{
A(i)\right\} _{\mathcal{I}}$ such that for every ball in every cover, the
sequence $\left\{ \int_{B_{n}(x_{n,m})}\left| F(i)\right| ^{2}dx\right\} _{%
\mathcal{I}^{\prime }}$ converges. \ For each $C_{n}$, consider the
collection of balls $\left\{ B_{n}(y_{n,m})\right\} $, $\left\{
y_{n,m}\right\} \subset \left\{ x_{n,m}\right\} $, for which $\left\{
\int_{B_{n}(y_{n,m})}\left| F(i)\right| ^{2}dx\right\} _{\mathcal{I}^{\prime
}}$ converges to a value greater than or equal to $\varepsilon $. \ Note
that for any $i\in \mathcal{I}$, there is an upper bound on the number $%
N_{i,n}$ of disjoint balls of radius $\frac{1}{n}$ for which $\int_{B_{n}(y_{n,m})}%
\left| F(i)\right| ^{2}dx\geq \varepsilon :$%
\begin{eqnarray*}
B\geq N_{i,n}\varepsilon.
\end{eqnarray*}%
Thus the upper bound $\frac{B}{\varepsilon }$ limits the number of
disjoint balls in the set $\left\{ B_{n}(y_{n,m})\right\} .$

Choose a maximal disjoint set $\left\{ B_{n}(y_{n,m_{j}})\right\} _{j=1}^{J}$
of balls in $\left\{ B_{n}(y_{n,m})\right\} $, and consider the set $\left\{
B_{n}^{\ast }(y_{n,m_{j}})\right\} _{j=1}^{J}$ of balls centered at the
points $y_{n,m_{j}}$ but having radius $\frac{3}{n}$. \ Then we have $%
\bigcup_{\left\{ y_{n,m}\right\} }B_{n}(y_{n,m})\subset
\bigcup_{j=1}^{J}B_{n}^{\ast }(y_{n,m_{j}}).$ \ This shows that if we
discard the balls $\left\{ B_{n}(y_{n,m})\right\} $ from the cover $C_{n}$,
we will only have discarded a set which was contained in $J\leq \frac{B}{%
\varepsilon }$ balls of radius $\frac{3}{n}$. \ We can then safely discard
the balls $\left\{ B_{n}(y_{n,m})\right\} $ from each cover $C_{n}$, and
form the union $C=\bigcup_{n\in \mathbb{N}}C_{n}\backslash \left\{
B_{n}(y_{n,m})\right\} $ to obtain a cover $C$ of $M\backslash \left\{
x_{1},...,x_{k}\right\} $, where $k\leq \frac{B}{\varepsilon }$, and each ball $B_{n}\left( x_{n,m} \right) \in C$ satisfies $\int_{B_{n}\left( x_{n,m} \right)} { \left| F \left( i \right) \right|^{2}}dx < \varepsilon$ .
\end{proof}

Since a minimizing sequence $\left\{ A(i)\right\} _{i\in \mathcal{I}}$ by
definition admits a uniform bound on the action, we can use Theorem \ref%
{goodcover} to select a subsequence $\left\{ A(i)\right\} _{i\in \mathcal{I}%
^{\prime }}$ of the minimizing sequence and a cover $\left\{ U_{\alpha
}\right\} $ satisfying $\int_{U_{\alpha }}\left| F(i)\right| ^{2}$ $dx<K(4)$ 
$\ \ \forall $ $i\in \mathcal{I}^{\prime }$, $i>I_{\alpha }$. \ On any
neighborhood $U_{\alpha }$ in the cover, Theorem \ref{localgauge} implies
that each member $A_{\alpha }(i)$ of the subsequence is gauge-equivalent to
a connection $\hat{A}_{\alpha }(i)$ in the Hodge gauge, satisfying a uniform 
$W^{1,2}(U)$ bound on $\hat{A}_{\alpha }(i)$. \ Weak compactness of Sobolev
spaces now yields a further subsequence of $\left\{ \hat{A}_{\alpha
}(i)\right\} $, weakly convergent in $W^{1,2}$ to some $\hat{A}_{\alpha }$.
\ It only remains to show that $\hat{A}_{\alpha }$ retains the desired\
regularity properties and boundary data, and that the set $\left\{ \hat{A}%
_{\alpha }\right\} $ can be patched to a global connection on $M.$ \ These
objectives are accomplished by Theorem 3.4 in \cite{Marini} (generalizing
Theorem 3.6 in \cite{Uhlenbeck} and Theorem 3.1 in \cite{Sedlacek}). \ Their
results are paraphrased below; the proof is by weak compactness of Sobolev
spaces:

\begin{theorem}
\label{goodlimit}
Let $\left\{ A(i)\right\} _{i\in \mathcal{I}}$ be a
sequence of $G$-connections with uniformly bounded action as described in
Theorem \ref{goodcover}, and with prescribed smooth tangential boundary
components $\left( A\left( i\right) \right) _{\tau }=a_{\tau }$ on $\partial
M$. \ Let $\varepsilon =K(4)$, where $K(4)$ is the constant from Theorem \ref%
{localgauge}. \ Then, for the subsequence $\left\{ A\left( i\right) \right\}
_{i\in \mathcal{I}^{\prime }}$ found in Theorem \ref{goodcover} and cover $%
\left\{ U_{\alpha }\right\} $, there exists a further subsequence $\left\{
A\left( i\right) \right\} _{i\in \mathcal{I}^{\prime \prime }}$, sections $%
\sigma _{\alpha }\left( i\right) :U_{\alpha }\rightarrow P_{i}$ ($i\in 
\mathcal{I}^{\prime \prime }$) and connections $A_{a}$ on $U_{\alpha }$ such
that \\ \\
\begin{tabular}{ll}
(e) & $\sigma _{\alpha }^{\ast }\left( i\right) \left( A_{i}\right) \equiv
A_{\alpha }\left( i\right) \rightharpoonup A_{\alpha }$ \ in $W^{1,2}\left(
U_{\alpha }\right) $ \\ 
(f) & $F\left( A_{\alpha }\left( i\right) \right) \equiv F_{\alpha }\left(
i\right) \rightharpoonup F_{\alpha }$ in $L^{2}\left( U_{\alpha }\right) $
\\ 
(g) & $s_{\alpha \beta }\left( i\right) \rightharpoonup s_{\alpha \beta }$
in $W^{2,2}\left( U_{\alpha }\cap U_{\beta }\right) $ \\ 
(h) & $\left. \left( A_{\alpha }\right) _{\tau } \right|
_{\partial _{1}U_{\alpha }}\sim \left. a_{\tau } \right|
_{\partial _{1}U_{\alpha }}$ by a smooth gauge transformation \\ 
(i) & $\left( 
\begin{array}{c}
d\ast A_{\alpha }=0\ on \ U_{\alpha } \\ 
d_{\tau }\ast \left( A_{\alpha }\right) _{\tau }=0\ on \ \partial _{1}U%
\end{array}%
\right) $ \\ 
(j) & $A_{\alpha }\equiv s_{\alpha \beta }^{-1}A_{\beta }s_{\alpha \beta
}+s_{\alpha \beta }^{-1}ds_{\alpha \beta }$%
\end{tabular}%
\\ \\
Here $s_{\alpha \beta }(i)$ is the transition function $A_{\beta }\left(
i\right) \rightarrow A_{\alpha }\left( i\right) $; i.e, 
\begin{eqnarray*}
A_{\alpha }\left( i\right) \equiv s_{\alpha \beta }^{-1}\left( i\right)
A_{\beta }\left( i\right) s_{\alpha \beta }\left( i\right) +s_{\alpha \beta
}^{-1}\left( i\right) ds_{\alpha \beta }\left( i\right) .
\end{eqnarray*}
\end{theorem}

\begin{proof}
See \cite{Sedlacek}, \cite{Marini}. \ Note that the result follows by weak
compactness of Sobolev spaces, after applying diagonalization (as in the
proof of Theorem \ref{goodcover}) over the countable cover $\left\{
U_{\alpha }\right\} $.
\end{proof}

Lower semicontinuity of the Yang-Mills functional now implies that the value
of the action on the limiting connection $A$ of the sequence described in
Theorem \ref{goodlimit} is in fact $m\left( a_{\tau }\right) \equiv
\min\limits_{\mathcal{A}}I\left( A\right) $ where $\mathcal{A}$ is the set
of connections on $G$-bundles on $M$ such that $\left. A_{\tau } \right| _{\partial M}$. \ In Theorem 3.5 in \cite{Marini} and 4.1 in \cite%
{Sedlacek}, it is proved by contradiction that $A$ in fact satisfies the
Yang-Mills equations. \ These proofs are completely local, and hold
unchanged in our case.

The proofs of regularity of the connection in Hodge gauge are also local and
hold unchanged. \ Regularity except for at the points $\left\{
x_{1},...,x_{k}\right\} $ from Theorem \ref{goodcover} is a consequence of
the ellipticity of the Yang-Mills equations in Hodge gauge. \ At the points $%
\left\{ x_{1},...,x_{k}\right\} $, the limiting connection may not be
defined, so removable singularity theorems are needed to extend $A$ to these
points. \ The case of interior points is covered by Theorem 4.1 in \cite%
{UhlRS}, and that of boundary points by Theorem 4.6 in \cite{Marini}, so that the connection $A$ extends to a smooth connection (provided the Dirichlet boundary data is
smooth). \ More precisely, we have

\begin{theorem}
\label{RS}
Let $U^{\left( 1\right) }$ ($U^{\left( 2\right) }$) be a
neighborhood of type 1 (2); let $U_{\ast }^{\left( i\right) }=U^{\left(
i\right) }\backslash \left\{ 0\right\} $. \ Let $A$ be a connection in a
bundle $P$ over $U_{\ast }^{\left( i\right) }$, $\left\| F_{A}\right\|
_{L^{2}\left( U\right) }<B<\infty $. \ Then%
\\
\begin{tabular}{ll}
(Type 1) & If $A$ is Yang-Mills on $\left. P\right| U_{\ast }^{\left(
1\right) }$, there exists a $C^{\infty }$ connection $A_{0}$ \\ 
& defined on $U^{\left( 1\right) }$ such that $A_{0}$ is gauge-equivalent to 
$A$ on $U_{\ast }^{\left( 1\right) }$. \\ 
(Type 2) & If $A$ is Yang-Mills and $C^{\infty }$ on $\left. P\right|
U_{\ast }^{\left( 2\right) }$, there exists a $C^{\infty }$ connec- \\ 
& tion $A_{0}$ defined on $U^{\left( 2\right) }$ such that $A_{0}$ is
gauge-equivalent to $A$ on \\ 
& $U_{\ast }^{\left( 2\right) }$, by a gauge transformation in $C^{\infty
}\left( U_{\ast }\right) $.%
\end{tabular}%
\end{theorem}

\begin{proof}
See \cite{UhlRS}, \cite{Marini}.
\end{proof}

\section{The Euclidean Yang-Mills Hamilton-Jacobi functional}
\label{Fun}

Having shown the existence of an absolute minimizer $A_{t}$ for the
Euclidean Yang-Mills action given prescribed smooth initial tangential
components $A=a_{\tau }$, we can now define the Hamilton-Jacobi functional%
\footnote{Note that for $S\left( A\right) $ to be finite, we are implicitly making the assumption that for all initial data $A$ of physical interest, there exists at least one trajectory $A_{s}$ ($A_{s=0}=A$) such that $-\tilde{I}\left(
A_{s}\right) <\infty $. \ This constraint defines the set of physical
fields, since for any $A$ on $\mathbb{R}^{3}$ for which no such $A_{s}$ can
be found, allowing $S\left( A\right) $ to take an infinite value implies
that evaluated on this $A$, the ground state $\Omega \left( A\right) $ is
zero.}%
\begin{eqnarray*}
S\left( A\right) =\int_{\mathbb{R}_{+}\times \mathbb{R}^{3}}tr\left(
F_{A_{t}}\wedge \ast F_{A_{t}}\right) dt,
\end{eqnarray*}%
where $\ast $ indicates the Hodge star operator in the Euclidean metric.

The values of this functional are well-defined even allowing for the
possible existence of more than one gauge-equivalence class of minimizers
for the given initial data; in principle we can simply choose a minimizer
starting from the field configuration $A=a_{\tau }.$ \ However, while still
an open question, there exist partial results toward establishing uniqueness
of a minimizer for given initial data in the compact case. \ In \cite{Isobe}%
, Isobe has shown that for flat boundary values, the Dirichlet problem on a
star-shaped bounded domain in $\mathbb{R}^{n}$\ can only have a flat
solution. \ Non-uniqueness results are proven by Isobe and Marini \cite%
{IsMar} for Yang-Mills connections in bundles over $B^{4}$, but the
solutions are topologically distinct, belonging to differing Chern classes.
\ On the domain $M=\mathbb{R}_{+}\times \mathbb{R}^{3}$, it seems
likely that given initial data determines a minimizer unique up to gauge
transformation. \ Future work will aim to settle this question; one possible
means of approach is a conformal transformation to the compact case, as
described in Sect.~\ref{YMDirichlet}.

In order to make the claim that $S(A)$ solves the imaginary-time zero-energy
Yang-Mills Hamilton-Jacobi equation, we must also verify its functional
differentiability. \ This can be done using the same integration by parts
argument as in the derivation of the Euler-Lagrange equation. \ However, we
must first write the solution to the Euclidean Dirichlet problem in a global
gauge which is smooth and decays sufficiently rapidly at spatial and
temporal infinity.

First, Theorem \ref{RS} implies that the solution $A$ to the Yang-Mills
Dirichlet problem extends to a smooth connection on a smooth bundle over all
of $M=\mathbb{R}_{+}\times \mathbb{R}^{3}$. \ Since the only bundle
over a contractible base manifold is the trivial one (see e.g. \cite{Nakahara}), $A$ is also a connection on the trivial bundle $P\cong M\times
G $. \ Therefore we can write $A$ in terms of a smooth global section $%
\sigma :M\rightarrow G$. \ Using this trivialization, $D=d+A$ is smoothly
defined over all of $M$.

The following lemma controls the growth of $A$ and $F,$ for a good choice of
gauge. \ Part \textit{(a)} is a version of Uhlenbeck's Corollary 4.2 \cite%
{UhlRS} for our base manifold $M=\mathbb{R}_{+}\times \mathbb{R}^{3}$;
part \textit{(b)} extends the same principle to bound the growth of the
connection 1-form $A$.

\begin{lemma}
\label{Decay}
Let $D=d+A$ be a connection in a bundle $P$ over an exterior
region $\mathcal{V}=\left\{ y\in \mathbb{R}_{+}\times \mathbb{R}%
^{3}:\left| y\right| \geq N\right\} $ satisfying $\int_{\mathcal{V}}\left|
F\right| ^{2}<\infty $. \ Then 
\\
\begin{tabular}{ll}
(a) & $\left| F\right| \leq C\left| y\right| ^{-4}$ for some constant $C$
(not uniform); \\ 
(b) & There exists a gauge in which $D=d+\tilde{A}$ satisfies $\left| 
\tilde{A}\right| \leq K\left| y\right| ^{-2}$.%
\end{tabular}%
\end{lemma}

\begin{proof}
\textit{(a)} \ Following the reasoning in \cite{UhlRS}, we define the
conformal mapping 
\begin{eqnarray*}
f:U_{\ast }\rightarrow \mathcal{V} \\ 
y=f\left( x\right) =N\frac{x}{\left| x\right| ^{2}},%
\end{eqnarray*}%
where $U_{\ast }=\left\{ x\in \mathbb{R}_{+}\times \mathbb{R}%
^{3}:0<\left| x\right| \leq 1\right\} $. \ By conformal invariance of the
Yang-Mills action, we have%
\begin{eqnarray*}
\int_{U_{\ast }}\left| f^{\ast }F\right| ^{2}=\int_{U_{\ast }}\left| F\left(
f^{\ast }D\right) \right| ^{2}=\int_{\mathcal{V}}\left| F\right| ^{2}.
\end{eqnarray*}%
Applying part (b) of Theorem \ref{RS} to the pullback $f^{\ast }D$ of $D$
under $f$, there exists a gauge transformation $\sigma :U_{\ast }\rightarrow
G$ in which $f^{\ast }D$ extends smoothly to $U$. \ Thus using the
transformation law for $2$-forms, we have the following 
\begin{eqnarray*}
\left| F\left( y\right) \right| &=&\left| f^{\ast }F(x)\right| \left|
df\left( x\right) \right| ^{-2} \\
&\leq &\max_{x\in U}\left| f^{\ast }F(x)\right| \cdot \left( N/\left|
x\right| ^{2}\right) ^{-2} \\
&=&C^{\prime }N^{2}\left| y\right| ^{-4}
\end{eqnarray*}

\textit{(b)} \ Define the gauge transformation $s=\sigma \circ f^{-1}:%
\mathcal{V}\rightarrow G$. \ Denoting $A^{s}=s^{-1}ds+s^{-1}As$ by $\tilde{A}
$ and $\left( f^{\ast }A\right) ^{\sigma }=\sigma ^{-1}d\sigma +\sigma
^{-1}\left( f^{\ast }A\right) \sigma $ by $\widetilde{f^{\ast }A}$, we have $%
f^{\ast }\tilde{A}=\widetilde{f^{\ast }A}$. \ Thus again applying Theorem
11(b) and using the transformation law for $1$-forms,%
\begin{eqnarray*}
\left| \tilde{A}\left( y\right) \right| &=&\left| f^{\ast }\tilde{A}%
(x)\right| \left| df\left( x\right) \right| ^{-1} \\
&\leq &\max_{x\in U}\left| \widetilde{f^{\ast }A}(x)\right| \cdot \left(
N/\left| x\right| ^{2}\right) ^{-1} \\
&=&C^{\prime \prime }N\left| y\right| ^{-2}.
\end{eqnarray*}%
\end{proof}

We are now ready to prove differentiability of our Hamilton-Jacobi
functional. \ Thanks are due to V. Moncrief for suggesting the form of this
argument.

\begin{theorem}
The functional%
\begin{eqnarray*}
S\left( A\right) =-\tilde{I}\left( A\right) =\int_{\mathbb{R}%
_{+}\times \mathbb{R}^{3}}tr\left( F_{A_{t}}\wedge \ast F_{A_{t}}\right) dt
\end{eqnarray*}%
is functionally differentiable, and $\frac{\delta S}{\delta A}=E=\dot{A}_{t=0}$.
\end{theorem}

\begin{proof}
To find the functional derivative of $S(A)=-\tilde{I}\left( A_{t}\right) $
at a given connection $A_{0}$ on the slice $x^{0}=0$, consider the
1-parameter family $A_{0}+\lambda h$, constructing%
\begin{eqnarray*}
\left. \frac{d}{d\lambda }\left[ S\left( A_{0}+\lambda h\right) \right]
\right| _{\lambda =0} &=&\lim\limits_{\lambda \rightarrow 0}\frac{S\left(
A_{\lambda }\right) -S(A_{0})}{\lambda } \\
&=&\lim\limits_{\lambda \rightarrow 0}\frac{-\tilde{I}\left( A_{\lambda
,t}\right) -\left( -\tilde{I}(A_{0,t})\right) }{\lambda } \\
&=&-\lim\limits_{\lambda \rightarrow 0}\frac{1}{\lambda }\left[ \tilde{I}%
\left( A_{\lambda ,t}\right) -\tilde{I}(A_{0,t})\right]
\end{eqnarray*}%
where for each $A_{\lambda }=A_{0}+\lambda h$, $A_{\lambda ,t}$ denotes the
absolute minimizer of $-\tilde{I}$ given initial data $A_{\lambda }$. \ For
any given value $\lambda _{0}$, the difference $\tilde{I}\left( A_{\lambda
_{0},t}\right) -\tilde{I}(A_{0,t})$ can be expressed in terms of a Taylor
series, as follows. \ First, use the parameter $\lambda $ to interpolate
between $A_{0,t}$ and $A_{\lambda _{0},t}$, describing a 1-parameter family $%
X_{\lambda ,t}$,%
\begin{eqnarray*}
X_{\lambda ,t}\equiv \frac{\lambda }{\lambda _{0}}A_{\lambda _{0},t}+\left(
1-\frac{\lambda }{\lambda _{0}}\right) A_{0,t},
\end{eqnarray*}%
so that $X_{\lambda ,0}=A_{\lambda }$. \ The standard Taylor series
expansion of $\tilde{I}\left( X_{\lambda ,t}\right) $ as a function of $%
\lambda $ then gives%
\begin{equation}
\tilde{I}\left( A_{\lambda _{0},t}\right) -\tilde{I}(A_{0,t})=\lambda
_{0}\left( \frac{\partial \tilde{I}}{\partial \lambda }\right) _{\lambda
=0}+O\left( \lambda _{0}^{2}\right) .  
\label{Idiff}
\end{equation}%
Let $h_{t}=\frac{1}{\lambda _{0}}\left( A_{\lambda _{0},t}-A_{0,t}\right) $,
so that $X_{\lambda ,t}=A_{0,t}+\lambda h_{t}$. \ Then%
\begin{eqnarray*}
\left. \frac{\partial \tilde{I}}{\partial \lambda }\right| _{\lambda =0}
&=&\left. \frac{\partial }{\partial \lambda }\left[ \int_{\mathbb{R}%
_{+}\times \mathbb{R}^{3}}\left\langle F_{X_{\lambda ,t}},F_{X_{\lambda
,t}}\right\rangle \right] \right| _{\lambda =0} \\
&=&2\int_{\mathbb{R}_{+}\times \mathbb{R}^{3}}\left\langle dh_{t}+%
\left[ A_{0,t},h_{t}\right] ,F_{A_{0,t}}\right\rangle \\
&=&2\lim_{R\rightarrow \infty }\left( \int_{\partial _{1}}\left\langle
h,F_{A_{0}}\right\rangle +\int_{\partial _{2}}\left\langle
h_{t},F_{A_{0,t}}\right\rangle -\int_{0\leq \left| x\right| <R}\left\langle
h_{t},D^{\ast }F_{A_{0,t}}\right\rangle \right)
\end{eqnarray*}%
where $\partial _{1}=\left\{ \left| x\right| <R,x^{0}=0\right\} ,$ $\partial
_{2}=\left\{ \left| x\right| =R,x^{0}>0\right\} $.

The last term on the right-hand side vanishes due to the fact that $%
F_{A_{0,t}}$ is a solution to the Yang-Mills equations. \ Working with $%
A_{\lambda _{0},t}$ and $A_{0,t}$ both in the gauge guaranteed by Lemma \ref%
{Decay} (for some fixed $N$ which $R$ eventually surpasses), the middle term
also approaches zero as $R$ approaches infinity, since%
\begin{eqnarray*}
\left\langle h_{t},F_{A_{0,t}}\right\rangle  &\leq &\left| h_{t}\right|
\left| F_{A_{0,t}}\right|  \\
&\leq &\frac{1}{\lambda _{0}}\left( \left| A_{\lambda _{0},t}\right| +\left|
A_{0,t}\right| \right) \left| F_{A_{0,t}}\right|  \\
&\leq &\frac{1}{\lambda _{0}}\left( K_{\lambda _{0}}+K_{0}\right) C_{0}\cdot
R^{-6}.
\end{eqnarray*}%
Since the area element on $\partial _{2}$ contributes only a factor of $R^{2}
$, the middle term is easily seen to vanish. \ Thus we are left with only
the first term, so that%
\begin{eqnarray*}
\left. \frac{\partial \tilde{I}}{\partial \lambda }\right| _{\lambda
=0}=\int_{\mathbb{R}^{3}}\left\langle h,F_{A_{0}}\right\rangle ,
\end{eqnarray*}%
and the definition of functional derivative implies that%
\begin{eqnarray*}
\frac{\delta S}{\delta A}=E=\dot{A}_{t=0}.
\end{eqnarray*}
\end{proof}

\section{Gauge and Poincare invariance}
\label{Gauge}

In order for the candidate ground state wave functional%
\begin{eqnarray*}
\Omega (A)= \mathcal{N} \exp \left( -S\left( A\right) \right) 
\end{eqnarray*}%
to be physical, it must remain invariant under the action of gauge
transformations $g\left( x\right) $, $x\in \mathbb{R}^{3}$, on the
connection $A(x):$%
\begin{eqnarray*}
S\left( g^{-1}dg+g^{-1}Ag\right) =S\left( A\right) ,
\end{eqnarray*}%
so that $S$ is in fact a functional on the physical configuration space $%
\mathcal{A/G}$ of connections modulo gauge transformations, rather than the
kinematical configuration space $\mathcal{A}$. \ Gauge invariance of $S$
follows immediately from its form%
\begin{eqnarray*}
S(A)=-\int_{0}^{\infty }\tilde{L}\left( A_{t},\dot{A}_{t}\right) dt=\int_{\mathbb{R}_{+}\times \mathbb{R}^{3}}tr\left( F_{A_{t}}\wedge
\ast F_{A_{t}}\right) 
\end{eqnarray*}%
where $\ast $ denotes the Hodge star operator in the Euclidean metric on $%
\mathbb{R}_{+}\times \mathbb{R}^{3}$. \ The gauge transformation $%
g\left( x\right) $, $x\in \mathbb{R}^{3}$, can simply be extended to $\mathbb{R}_{+}\times \mathbb{R}^{3}$ by taking $g\left( t,x\right) =g\left(
x\right) $ constant over $\mathbb{R}_{+}$, and the cyclic property of
the trace implies%
\begin{eqnarray*}
S\left( g^{-1}dg+g^{-1}Ag\right)  &=&\int_{\mathbb{R}_{+}\times 
\mathbb{R}^{3}}tr\left( F_{g\cdot A_{t}}\wedge \ast F_{g\cdot A_{t}}\right) 
\\
&=&\int_{\mathbb{R}_{+}\times \mathbb{R}^{3}}tr\left( F_{A_{t}}\wedge
\ast F_{A_{t}}\right) =S\left( A\right) .
\end{eqnarray*}

Similarly, rotations and translations applied to $\mathbb{R}^{3}$ do not
affect the value of $S\left( A\right) $, because we can extend them
constantly through time over $\mathbb{R}_{+}\times \mathbb{R}^{3}$,
and by a change of coordinates the value of the integral defining $S\left(
A\right) $ is unchanged. \ The only remaining Poincare transformations are
boosts, which cannot be verified directly in our canonical framework. \ The
conserved quantity generating an infinitesimal boost in the $x^{i}$
direction is 
\begin{equation}
C_{B(i)}=\int_{\mathbb{R}^{3}}\left( x^{0}\delta^{i}_{\mu} + x^{i}\delta^{0}_{\mu} \right) T^{\mu 0}d^{3}x,  
\label{boost}
\end{equation}%
where $T^{\mu \nu }=-\frac{1}{4\pi } tr \left\{ F_{\  \ \alpha }^{\mu
}F^{\nu \alpha }-\frac{1}{4}\eta ^{\mu \nu }F_{\alpha \beta }F^{\alpha \beta
}\right\} $ is the stress-energy tensor of Yang-Mills theory. \ This
infinitesimal generator must be promoted to a quantum operator which annihilates our candidate ground state. \ A test case in which this can be done is the
abelian case of $U\left( 1\right) $ gauge theory (free Maxwell theory),
using Wheeler's zero-energy ground state (\ref{Wheeler}) as in Sect.~\ref{Intro}%
. \ This is alternatively attainable as%
\begin{eqnarray*}
\Omega (A)=\mathcal{N}\exp \left( -\frac{\left\langle A, \left( \ast d \right)\triangle ^{-1/2}\left( \ast d\right) A\right\rangle _{2}}{2%
}\right) 
\end{eqnarray*}%
by using the normal ordering as described in Sect.~\ref{Ordering} to write the
unique positive square root of the operator $\ast d\ast d$ in the form $%
\mathbf{(}\ast d)\triangle ^{-1/2}\left( \ast d\right) $. \ Writing the
operator $\triangle ^{-1/2}$ in terms of its integral kernel shows the
equality of this state with (\ref{Wheeler}). \ Invariance under
infinitesimal boosts follows by expressing (\ref{boost}) as the sum of a
translation generator $\frac{x^{0}}{4\pi }\int_{\Sigma }\varepsilon
^{ijk}E_{j}B_{k}d^{3}x$ plus the term $\frac{1}{8\pi }\int_{\Sigma
}x^{i}\left( \left| E\right| ^{2}+\left| B\right| ^{2}\right) d^{3}x$. \
Translation invariance being already established, it only remains to verify
that (\ref{Wheeler}) is annihilated by the remaining term under our
ordering. \ Indeed, the functional $S(A)$ in the exponent of (\ref{Wheeler})
can be directly shown to satisfy%
\begin{eqnarray*}
\int_{\mathbb{R}^{3}}x^{i}\left| \frac{\delta S}{\delta A}\right|
^{2}d^{3}x=\int_{\mathbb{R}^{3}}x^{i}\left| B\right| ^{2}d^{3}x,
\end{eqnarray*}%
allowing the extra term to be ordered in the same way as the Hamiltonian. \
Using the abelian case as a model, we hope to extend invariance under boosts
to the nonabelian case in future work.
\\ \\
\textit{ \textbf{Acknowledgements.} I am very grateful to my advisor, Vincent Moncrief, for suggesting this project, and for help and guidance throughout its development. \ For kind hospitality, I offer my heartfelt thanks to the University of Amsterdam's
Korteweg-de Vries Institute for Mathematics, in particular Eric Opdam and
Jan Wiegerinck.}

\pagebreak

\end{document}